\documentclass[letterpaper,preprintnumbers,prd,twocolumn,nofootinbib,nobibnotes,showpacs]{revtex4}
\usepackage{graphics,epsfig,subfigure}
\usepackage{color}
\usepackage{amsfonts}
\usepackage{mathrsfs}
\usepackage{epsfig}
\usepackage{graphicx}%
\usepackage{dcolumn}
\usepackage{amsmath}
\usepackage{color}
\usepackage{color}

\begin{document}
\renewcommand{\baselinestretch}{1.3}
\newcommand\beq{\begin{equation}}
\newcommand\eeq{\end{equation}}
\newcommand\beqn{\begin{eqnarray}}
\newcommand\eeqn{\end{eqnarray}}
\newcommand\nn{\nonumber}
\newcommand\fc{\frac}
\newcommand\lt{\left}
\newcommand\rt{\right}
\newcommand\pt{\partial}

\allowdisplaybreaks

\title{Black Holes With Many Horizons in the Theories of Nonlinear Electrodynamics}
\author{Changjun Gao\footnote{gaocj@nao.cas.cn}}

\affiliation{National Astronomical Observatories, Chinese Academy of Sciences, 20A Datun Road, Beijing 100101, China}

\affiliation{School of Astronomy and Space Sciences, University of Chinese Academy of Sciences,
19A Yuquan Road, Beijing 100049, China}

\begin{abstract}
In this article, we construct exact black hole solutions with many horizons (more than two) in the Einstein-nonlinear electrodynamic theories. In particular, we acquire the explicit expression of nonlinear electrodynamic Lagrangian for the 3-horizon black holes. Then we make the investigations of 3-horizon black holes on the horizons, the null and timelike geodesics, the Love numbers and the thermodynamics.
\end{abstract}











\maketitle



\section{Introduction}
In order to remove the inconsistencies of infinite self-energy of the point charges in the Maxwell's electrodynamics,
Max Born firstly proposed the nonlinear electrodynamic (NLE) theory \cite{bor:1934}. The theory was subsequently extended in collaboration
with Leopard Infield \cite{inf:1934} to the so-called Born-Infield theory. It is amazing that the Born-Infield theory was rediscovered
in the low energy limit of the string theory \cite{fra:1985} half a century later. It is found that the Born-Infield parameter $b$ is related to the string tension $\alpha^{'}$ via $2\pi\alpha^{'} = 1/b$ \cite{tse:1999}. On the other hand, not long after
the proposal of Born and Infield, Warner K. Heisenberg and Khans H. Euler \cite{hei:1936} achieved the one-loop correction of
quantum electrodynamics to Maxwell's Lagrangian.

Take into account gravity, one suppose the NLE theories may erase the black hole singularities.
However, it is not the case. It is found neither electrically charged Einstein-Born-Infield black holes \cite{dey:2004,fer:2003,gar:1984,sal:1987} nor electrically charged Einstein-Euler-Heisenberg black holes \cite{ruf:2013,yaj:2001} are regular.
The first and remarkable regular black hole solution was written down by Bardeen \cite{bar:1968} by hand. It was interpreted very latter by Ayon-Beat and
Garcia \cite{ayo:1998, ayo:2000} as the solution of a particular Einstein-NLE theory.
Later on, a lot of regular black hole solutions were found within Einstein-NLE theories based
on various Lagrangian functions, such as logarithmic \cite{sol:1995}, hyperbolic tangent \cite{ayo:1999}, power
\cite{has:2007,has:2008}, exponential \cite{hen:2013}, de Sitter or anti-de Sitter asymptotic \cite{fan:2016} and so on. Finally, the black holes with charged scalar hairs in Einstein-NLE is considered in \cite{an:2021}. In the respect of regular black holes in NLE, an important point is the
existence of a no-go theorem \cite{Bron:1976, Bron:1979}. It is shown that there is no
such Lagrangian function which has the Maxwell weak field
limit that the resulting black hole has the regular center. In order to circumvent the no-go theorem, an interesting proposal is to consider a kind of phase transition on a certain sphere, outside of which there is a pure electric field but inside of which
the field is pure magnetic \cite{bur:2002}.

On the other hand, it is well-known that the Schwarzschild black hole has one event horizon. The Reissner-Nordstrom (RN) black hole, the Kerr black hole and the Kerr-Newman black hole have two horizons among which one is the event horizon and the other is the inner Cauchy horizon. So one interesting question that one would ask is `` Can a black hole have horizons more than two? '' The answer is yes. In fact, regular and non-regular multi-horizon (more than two horizons) black holes in Einstein-NLE theories have been presented in Ref.~\cite{noj:2017,gao:2018}.

\emph{It is found that the multi-horizon black holes have very rich physics. For instance, different from
the standard one-horizon black holes, the multi-horizon black holes show up not only the Hawking evaporation but
also anti-evaporation (or related instability phenomenon). In detail, one can consider the limit where the radius of one horizon coincides with that of another horizon for multi-horizon black holes. This is called the Nariai limit. Usually, the radius of the horizon decreases by the Hawking
radiation. However, in case of the Nariai limit, the radius can increase due to
the quantum effects. This is the well-known anti-evaporation effect \cite{bou:1998,noj:1999a,noj:1999b}. }

However, to the best of our knowledge, none of th solutions in Ref.~\cite{noj:2017,gao:2018} brings us with an analytic expression for NLE Lagrangian. In this article, starting from the most general, analytic form of Lagrangian (the sum of infinite series of Maxwell invariant) for NLE, we shall seek for the multi-horizon black hole solutions. In particular, we find and investigate the solution for a 3-horizon black hole with explicit form of NLE Lagrangian.

{The paper is organized as follows. In Sec. II, we report that we find the solutions for black holes with many horizons, in particular with three horizons, in NLE. In Sec. III, we give a detailed analysis on the horizons for the 3-horizon black holes. In Sec. IV, we derive the geodesic equation in the equatorial plane for both null and timelike geodesics in the background of 3-horizon black hole. Sec. V and Sec. VI are devoted to the investigations on radial geodesics, general geodesics, stable circular orbits and innermost stable circular orbit (ISCO). In Sec. VII, we consider the motion of test charged particles in the background of 3-horizon black hole. In Sec. VIII, we show the Love number of 3-horizon black hole is vanishing. This reveals the 3-horizon black hole is totally rigid. In Sec. IX, we make a study on the thermodynamics for 3-horizon black holes. Finally, we give the conclusion and discussion in Sec. X. Throughout this paper, we adopt the system of units in which $G=c=\hbar=1$ and the metric signature $(-,\ +,\ +,\ +)$.}

\section{solutions for black holes with many horizons}
We consider the Einstein theory coupled with the nonlinear electromagnetic field which has the action
\begin{equation}
S=\int d^4x\sqrt{-g}\left(R+L_{EM}\right)\;,
\end{equation}
with
\begin{eqnarray}
L_{EM}&\equiv& \sum_{i=1}^{\infty}\alpha_i \left(F^2\right)^{i}\;,\ \ \ \  F^2\equiv F_{\mu\nu}F^{\mu\nu}\;,\nonumber\\ F_{\mu\nu}&\equiv&\nabla_{\mu}A_{\nu}-\nabla_{\nu}A_{\mu}\;.
\end{eqnarray}
Here $R$ is the Ricci scalar and $L_{EM}$ is the extended Maxwell Lagrangian. $A_{\mu}$ is the Maxwell field. $\alpha_i$ are dimensional constants and have the dimension of ${\emph{lengh}}^{2(i-1)}$. $i$ is a positive integer. \emph{The physical motivation for taking the extended Maxwell Lagrangian Eq.~(2) for the
gauge field  is that it can cover nearly all the known proposals for NLE, for example, the Born-Infield-Lagrangian \cite{bor:1934,inf:1934}, the Euler-Heisenberg
Lagrangian \cite{hei:1936}, the power-law Maxwell Lagrangian \cite {has:2007,has:2008} and so on}.
When $\alpha_1=1$ and $\alpha_i=0$ (for $i>1$), it reduces to the Einstein-Maxwell theory. The variation of the action with respect to the metric gives
the Einstein equations
\begin{eqnarray}
G_{\mu\nu}&=&-2L_{EM,F^2}F_{\mu\lambda}F_{\nu}^{\lambda}+\frac{1}{2}g_{\mu\nu}L_{EM}\;, \nonumber\\ L_{EM,F^2}&\equiv&\frac{dL_{EM}}{dF^2}\;.
\end{eqnarray}
On the other hand, the variation of the action with respect to the field $A_{\mu}$ gives the generalized
Maxwell equations
\begin{eqnarray}
\nabla_{\mu}\left(L_{EM,F^2}F^{\mu\nu}\right)=0\;.
\end{eqnarray}
We shall look for the static, spherically symmetric black hole solutions in the theory. To this end, we take the ansatz of the metric
\begin{equation}
ds^2=-U\left(r\right)dt^2+\frac{1}{U\left(r\right)}dr^2+f\left(r\right)^2d\Omega_2^2\;,
\end{equation}
and the Maxwell field
\begin{equation}
A_{\mu}=\left[\Phi\left(r\right),\ \ 0,\ \ 0,\ \ 0\right]\;.
\end{equation}
Then the Einstein and Maxwell equations give
\begin{eqnarray}
&&f^{''}=0\;,\label{7} \\
&&U^{''}f+2U^{'}f^{'}+2Uf^{''}\nonumber\\&&+f\sum_{i=1}^{\infty}\left(-1\right)^{i}2^{i}\alpha_i \left(\Phi^{'}\right)^{2i}=0\;,\label{8}\\
&&f^2\sum_{i=1}^{\infty}i\left(-1\right)^{i}2^{i-1}\alpha_i \left(\Phi^{'}\right)^{2i-1}-Q=0\;,\label{9}\\
&&1-ff^{'}U^{'}-Uf^{'2}\nonumber\\&&+f^2\sum_{i=1}^{\infty}\left(-1\right)^{i}\left(2i-1\right)2^{i-1}\alpha_i \left(\Phi^{'}\right)^{2i}=0\label{10}\;.
\end{eqnarray}
Here $Q$, as an integration constant, is nothing but the electric charge of the RN black hole. The prime denotes the derivative with respect to $r$.
Solving Eq.~(\ref{7}), we obtain
\begin{eqnarray}\label{11}
f&=&r\;.
\end{eqnarray}
It seems rather difficult to solve remaining equations. But in fact, the solutions with electric charge have been solved in a general form by Pellicer and Torrence in \cite{pel:1969}. The magnetic counterpart was solved by Bronnikov in \cite{bron:2001}. Here we will look for the solutions in the form of series (the most general form). So we expand $U$ and $\Phi$ as follows
\begin{eqnarray}\label{12}
\Phi=\sum_{i=1}^{\infty}b_i r^{-i}\;,\ \ \ \ \ \ \ U=1+\sum_{i=1}^{\infty}c_i r^{-i}\;,
\end{eqnarray}
such that $\Phi$ is asymptotically vanishing and the spacetime is asymptotically Minkowski. Here $b_i$ and $c_i$ are constants. Taking Eq.~(\ref{11}) into account and substituting Eqs.~(\ref{12}) into Eqs.~(\ref{8}-\ref{10}), we obtain the non-vanishing constants when $\alpha_1=1$
\begin{eqnarray}
b_1&=&Q\;,\\
b_5&=&\frac{4}{5}Q^3\alpha_2\;,\\
b_9&=&\frac{4}{3}Q^5\left(4\alpha_2^2-\alpha_3\right)\;,\\
b_{13}&=&\frac{32}{13}Q^7\left(24\alpha_2^3-12\alpha_3\alpha_2+\alpha_4\right)\;,\\
b_{17}&=&\frac{80}{17}Q^9\left(176\alpha_2^4-132\alpha_2^2\alpha_3+16\alpha_4\alpha_2\right.\nonumber\\&&\left.+9\alpha_3^2-\alpha_5\right)\;,\\
b_{21}&=&\frac{64}{7}Q^{11}\left(1456\alpha_2^5+234\alpha_3^2\alpha_2+208\alpha_4\alpha_2^2-24\alpha_4\alpha_3\right.\nonumber\\&&\left.-1456\alpha_2^3\alpha_3-20\alpha_5\alpha_2+\alpha_6\right)\;,\\
b_{25}&=&\frac{448}{25}Q^{13}\left(13056\alpha_2^6+2560\alpha_2^3\alpha_4-720\alpha_3\alpha_2\alpha_4\right.\nonumber\\&&\left.
+16\alpha_4^2-300\alpha_2^2\alpha_5+4320\alpha_3^2\alpha_2^2
-16320\alpha_2^4\alpha_3\right.\nonumber\\&&\left.+24\alpha_6\alpha_2-135\alpha_3^3+30\alpha_3\alpha_5-\alpha_7\right)\;,\\
\cdot\cdot\cdot&=&\cdot\cdot\cdot\cdot\cdot\cdot\;,
\end{eqnarray}
and
\begin{eqnarray}
c_1=c_1\;,\ \ \ \ \ \ \ c_i=\frac{4Q}{i+2}b_{i-1}\;,\ \ \  \textrm{for }\ \ \ \ i>1\;.
\end{eqnarray}
These form a solution of power series. For this solution, there are only two integration constants, $Q$ and $c_1$. $Q$ is the electric charge and $c_1=-2M$ ($M$ is the mass of the black hole). Basing on this solution, we have the following conclusions.
\subsection{2-horizon black hole}
When
\begin{eqnarray}
\alpha_1=1\;,\ \ \  \alpha_i=0\;,\ \ \ \left(i>1\right)\;,
\end{eqnarray}
we obtain
\begin{eqnarray}
\Phi=\frac{Q}{r}\;,\ \ \  U=1+\frac{c_1}{r}+\frac{Q^2}{r^2}\;.
\end{eqnarray}
Let $c_1=-2M$, then it is the RN black hole. $M$ and $Q$ represent the mass and electric charge of the black hole, respectively. In general, RN spacetime has two horizons. The corresponding Lagrangian $L_{EM}$ is given by
 \begin{eqnarray}
L_{EM}=F^2\;.
\end{eqnarray}
\subsection{3-horizon black hole}
\begin{figure}[htbp]
	\centering
	\includegraphics[width=8cm,height=6cm]{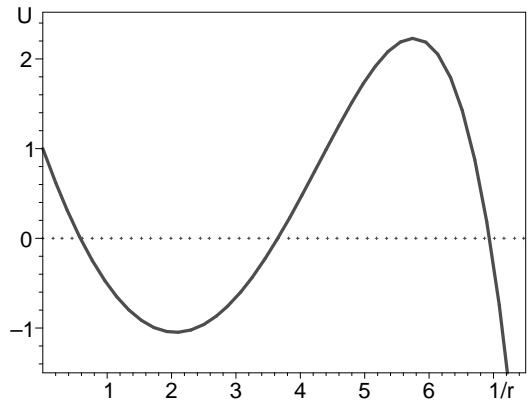}
	\caption{The locations of three horizons for $M=1,\ \ Q=0.7,\ \ \ \alpha_2=-0.001$.}
	\label{three}
\end{figure}

When
\begin{eqnarray}\label{co-3bh}
&&\alpha_1=1\;,\ \ \ \ \
\alpha_2\neq 0\;,\ \ \ \ \ \
\nonumber\\ &&\alpha_3=4\alpha_2^2\Longleftrightarrow b_9=0\;,\ \ \ \ \ \ \ \
\nonumber\\ &&\alpha_4=24\alpha_2^3\Longleftrightarrow b_{13}=0\;,\ \ \ \
\nonumber\\ &&\alpha_5=176\alpha_2^4\Longleftrightarrow b_{17}=0\;,\ \ \ \
\nonumber\\ &&\alpha_6=1456\alpha_2^5\Longleftrightarrow b_{21}=0\;,\ \ \ \
\cdot\cdot\cdot\;.
\end{eqnarray}
we obtain the solution
\begin{eqnarray}
\Phi&=&\frac{Q}{r}+\frac{4\alpha_2 Q^3}{5}\cdot\frac{1}{r^5}\;,\ \ \  \ \ \nonumber\\ U&=&1-\frac{2M}{r}+\frac{Q^2}{r^2}+\frac{2\alpha_2 Q^4}{5}\cdot\frac{1}{r^6}\;.
\end{eqnarray}
Given the expressions of $\Phi$, $U$ above and using the Lagrangian-generation-method in \cite{bea:1998}, we can construct the corresponding Lagrangian $L_{EM}$
\begin{eqnarray}\label{LM}
L_{EM}=\frac{1}{12\alpha_2}\left[1-\left(\zeta^{1/3}+\zeta^{-1/3}-1\right)^2\right]\;,
\end{eqnarray}
with
\begin{eqnarray}\label{LEM0}
\zeta\equiv 1-27\alpha_2F^2+3\sqrt{-6\alpha_2F^2+81\alpha_2^2\left(F^2\right)^2}\;.
\end{eqnarray}
We note that the Lagrangian is dependent on not the mass $M$ and charge $Q$, but only the coupling constant $\alpha_2$.
Expand  Eq.~(\ref{LM}) in series of $F^2$, we obtain
\begin{eqnarray}
L_{EM}&=&F^2+\alpha_2\left(F^2\right)^2+4\alpha_2^2\left(F^2\right)^3+24\alpha_2^3\left(F^2\right)^4\nonumber\\&&
+176\alpha_2^4\left(F^{2}\right)^5+1456\alpha_2^5\left(F^{2}\right)^6+\cdot\cdot\cdot\;,
\end{eqnarray}
which are consistent with the coefficients in Eq.~(\ref{co-3bh}).
In order that the root in Eq.~(\ref{LEM0}) always makes a sense, we should require
\begin{eqnarray}
\alpha_2\geq 0\;,\ \ \ \ \textrm{or}\ \ \ \ \alpha_2\leq \frac{2}{27F^2}\;,
\end{eqnarray}
 because of $F^2=-2\Phi^{'2}<0$. We shall find shortly later this spacetime can have three horizons when $\alpha_2$ is negative. As an example, we plot the locations of three horizons for $M=1,\ \ Q=0.7,\ \ \ \alpha_2=-0.001$ in Fig.~\ref{three}.\emph{ We point that there is no physical reasons for the
choice of above parameters. We simply choose the parameters by hand. }
\subsection{4-horizon black hole}
When
\begin{eqnarray}
&&\alpha_1=1\;,\ \ \ \ \
\alpha_2\neq 0\;,\ \ \ \ \
\alpha_3\neq 0\;,\ \ \ \ \
\nonumber \\ &&\alpha_4=12\alpha_2\left(-2\alpha_2^2+\alpha_3\right)\Longleftrightarrow b_{13}=0\;,\ \ \ \ \ \nonumber\\&&
\nonumber\\ &&\alpha_5=-208\alpha_2^4+60\alpha_2^2\alpha_3+9\alpha_3^2\Longleftrightarrow b_{17}=0\;,\ \ \ \ \ \nonumber\\&&
\nonumber\\ &&\alpha_6=26\alpha_2\left(9\alpha_3^2-24\alpha_2^4-16\alpha_2^2\alpha_3\right)\Longleftrightarrow b_{21}=0\;,\nonumber\\
&&\cdot\cdot\cdot\;.
\end{eqnarray}

we obtain the solution
\begin{eqnarray}
&&\Phi=\frac{Q}{r}+\frac{4\alpha_2 Q^3}{5}\cdot\frac{1}{r^5}+\frac{4\left(4\alpha_2^2-\alpha_3\right)Q^5}{3}\cdot\frac{1}{r^9}\;,\\ &&U=1-\frac{2M}{r}+\frac{Q^2}{r^2}+\frac{2\alpha_2 Q^4}{5r^6}+\frac{\left(16\alpha_2^2-4\alpha_3\right)Q^6}{9r^{10}}\;.
\end{eqnarray}

\begin{figure}[htbp]
	\centering
	\includegraphics[width=8cm,height=6cm]{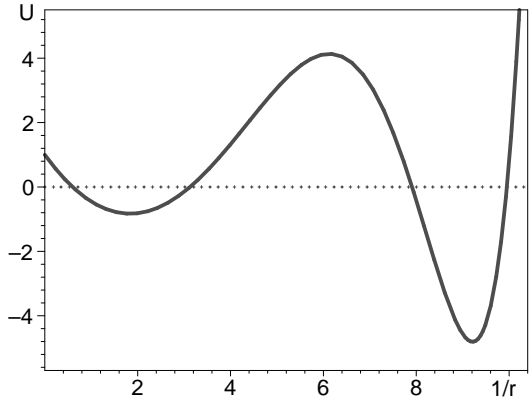}
	\caption{The locations of four horizons for $M=1,\ \ Q=0.74,\ \ \ \alpha_2=-0.0009,\ \ \ \alpha_3=3.14\cdot0^{-6}$.}
	\label{four}
\end{figure}

Using the Lagrangian-generation-method in \cite{bea:1998}, we find the corresponding Lagrangian $L_{EM}$ is

\begin{eqnarray}
L_{EM}=-2\eta^2-12\eta^4\alpha_2-40\left(4\alpha_2^2-\alpha_3\right)\eta^6\;,
\end{eqnarray}
where $\eta$ is determined by
\begin{eqnarray}
F^2=-2\left[\eta+4\alpha_2\eta^3+12\left(4\alpha_2^2-\alpha_3\right)\eta^5\right]^2\;.
\end{eqnarray}
Similarly, the Lagrangian is dependent on not the mass $M$ and charge $Q$, but only the coupling constants $\alpha_2$ and $\alpha_3$.
If the parameters, $M,\ Q,\ \alpha_2$ and $\alpha_3$ are properly chosen, the spacetime can have four horizons. As an example, we plot the locations of four horizons for $M=1,\ \ Q=0.74,\ \ \ \alpha_2=-0.0009,\ \ \ \alpha_3=3.14\cdot0^{-6}$ in Fig.~\ref{four}. \emph{ Same as the scenario of 3-horizon black holes, there is no physical reasons for the choice of above parameters. They are chosen simply by hand.}

Similarly, we can obtain black holes with 5-horizon, 6-horizon and so on. But in the next sections, we shall focus on 3-horizon black holes. Our lagrangians have intrinsics advantage over previous studies \cite{noj:2017,gao:2018}.  They are not dependent on the mass $M$, charge $Q$, or the ratio between charge to mass $Q/M$.

\section{horizons of 3-horizon black hole}
\begin{figure}[htbp]
	\centering
	\includegraphics[width=8cm,height=6cm]{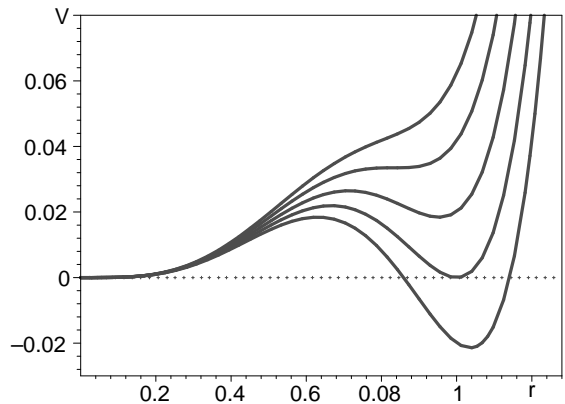}
	\caption{Five critical curves for $V$ with $Q^2>\frac{25}{24}M^2$, $Q^2=\frac{25}{24}M^2$, $M^2<Q^2<\frac{25}{24}M^2$, $Q^2=M^2$ and $Q^2<M^2$, from up to down, respectively.}
	\label{five}
\end{figure}

The positions of horizons for 3-horizon black holes are determined by
\begin{eqnarray}\label{eh}
U=0\;,\ \ \ \ \Longleftrightarrow \ \ \ \ \ V+\frac{2}{5}\alpha_2Q^4=0\;,
\end{eqnarray}
where $V$ is defined by
\begin{eqnarray}
V\equiv r^6-2Mr^5+Q^2r^4\;.
\end{eqnarray}
In order to determine the number of horizons, we make an analysis on $V$ in the first place. In Fig.~\ref{five}, we plot five critical curves for $V$ with $Q^2>\frac{25}{24}M^2$, $Q^2=\frac{25}{24}M^2$, $M^2<Q^2<\frac{25}{24}M^2$, $Q^2=M^2$ and $Q^2<M^2$, from up to down, respectively.
We note that $r=0$ is the curvature singularity. When $Q^2\geq\frac{25}{24}M^2$, $V$ is an increasing function. When $Q^2<\frac{25}{24}M^2$,
there are in general two local extremals, the local maximum $V_{max}=V\mid_{r=r_{-}}$ and local minimum $V_{min}=V\mid_{r=r_{+}}$ at
\begin{eqnarray}\label{r-}
r_{-}=\frac{5}{6}M\left(1-\sqrt{1-\frac{24Q^2}{25M^2}}\right)\;,
\end{eqnarray}
and
\begin{eqnarray}\label{r+}
r_{+}=\frac{5}{6}M\left(1+\sqrt{1-\frac{24Q^2}{25M^2}}\right)\;,
\end{eqnarray}
respectively. In particular, when $M^2<Q^2<\frac{25}{24}M^2$, the local minimum is positive. On the other hand, when $Q^2<M^2$, the local minimum is negative. Finally, when $Q^2=M^2$, the local minimum is zero.
Therefore, we have the following conclusions.
\subsection{When $Q^2\geq\frac{25}{24}M^2$}

In this case, $V$ is an increasing function of $r$. Therefore, if $\alpha_2<0$, there is only one black hole event horizon. But if $\alpha_2>0$, there is no horizon and the singularity is naked.

\subsection{When $M^2<Q^2<\frac{25}{24}M^2$}
In this case, we have $V_{max}>0$ and $V_{min}>0$. Thus we have the following conclusions.

1. If

\begin{eqnarray}
V_{max}+\frac{2}{5}\alpha_2Q^4<0\;,
\end{eqnarray}
there is only one black hole event horizon. We conclude $\alpha_2<0$.

2. If

\begin{eqnarray}
V_{max}+\frac{2}{5}\alpha_2Q^4=0\;,
\end{eqnarray}
there are two horizons. We conclude $\alpha_2<0$.

3. If

\begin{eqnarray}
0<V_{max}+\frac{2}{5}\alpha_2Q^4<V_{max}-V_{min}\;,
\end{eqnarray}
there are three horizons. We conclude $\alpha_2<0$.

4. If

\begin{eqnarray}
V_{min}+\frac{2}{5}\alpha_2Q^4=0\;,
\end{eqnarray}
there are two horizons. We conclude $\alpha_2<0$.

5. If

\begin{eqnarray}
V_{min}>V_{min}+\frac{2}{5}\alpha_2Q^4>0\;,
\end{eqnarray}
there is one horizon. We conclude $\alpha_2<0$.

6. If

\begin{eqnarray}
V_{min}+\frac{2}{5}\alpha_2Q^4\geq V_{min}\;,
\end{eqnarray}
there is no horizon. We conclude $\alpha_2\geq 0$.

\subsection{When $M^2=Q^2$}
In this case, we have $V_{max}>0$ and $V_{min}=0$.

1. If

\begin{eqnarray}
V_{max}+\frac{2}{5}\alpha_2Q^4<0\;,
\end{eqnarray}
there is only one black hole event horizon. We conclude $\alpha_2<0$.

2. If

\begin{eqnarray}
V_{max}+\frac{2}{5}\alpha_2Q^4=0\;,
\end{eqnarray}
there are two horizons. We conclude $\alpha_2<0$.

3. If

\begin{eqnarray}
0<V_{max}+\frac{2}{5}\alpha_2Q^4<V_{max}-V_{min}\;,
\end{eqnarray}
there are three horizons. We conclude $\alpha_2<0$.

4. If

\begin{eqnarray}
V_{min}+\frac{2}{5}\alpha_2Q^4=0\;,
\end{eqnarray}
there is one horizon. Since $V_{min}=0$, we conclude $\alpha_2=0$. It is nothing but the extreme Reissner-Nordstrom black hole.

5. If

\begin{eqnarray}
V_{min}+\frac{2}{5}\alpha_2Q^4>0\;,
\end{eqnarray}
there is no horizon. We conclude $\alpha_2>0$.

\subsection{When $Q^2<M^2$}

In this case, we have $V_{max}>0$ and $V_{min}<0$. Thus we have the following conclusions.

1. If

\begin{eqnarray}
V_{max}+\frac{2}{5}\alpha_2Q^4<0\;,
\end{eqnarray}
there is only one black hole event horizon.  We conclude $\alpha_2<0$.

2. If

\begin{eqnarray}
V_{max}+\frac{2}{5}\alpha_2Q^4=0\;,
\end{eqnarray}
there are two horizons.  We conclude $\alpha_2<0$.

3. If

\begin{eqnarray}
0<V_{max}+\frac{2}{5}\alpha_2Q^4<V_{max}\;,
\end{eqnarray}
there are three horizons. We conclude $\alpha_2<0$.

4. If

\begin{eqnarray}
V_{min}+\frac{2}{5}\alpha_2Q^4=0\;,
\end{eqnarray}
there is one horizon. We conclude $\alpha_2>0$.

5. If

\begin{eqnarray}
V_{min}<V_{min}+\frac{2}{5}\alpha_2Q^4<0\;,
\end{eqnarray}
there are two horizons. We conclude $\alpha_2>0$.

6. If

\begin{eqnarray}
V_{min}+\frac{2}{5}\alpha_2Q^4>0\;,
\end{eqnarray}
there is no horizon and the singularity is naked. We conclude $\alpha_2>0$. In summary, when $\alpha_2<0$, the spacetime can have three horizons. We shall mainly focus on $\alpha_2<0$ in the next sections.

\section{Equation for geodesics in the spacetime of 3-horizon black hole}
In this section, we give the equation of motion for null and timelike geodesics. In this section and later sections, we shall closely follow the way of Chandrasekhar \cite{chan:1985}. The equations governing the geodesics in a spacetime with the line element
\begin{eqnarray}
ds^2=g_{\mu\nu}{dX^{\mu}}{dX^{\nu}}\;,
\end{eqnarray}
can be derived from the Lagrangian
\begin{eqnarray}
\mathscr{L}
=\frac{1}{2}g_{\mu\nu}\frac{dX^{\mu}}{d\tau}\frac{dX^{\nu}}{d\tau}\;,
\end{eqnarray}
where $\tau$ is the some affine parameter along the geodesic. For timelike geodesics, $\tau$ can be identified with the proper time, $s$, of the co-moving observer along the geodesic.
For the 3-horizon black hole spacetime, the Lagrangian is
\begin{eqnarray}
\mathscr{L}
=\frac{1}{2}\left(U\dot{t}^{2}-\frac{1}{U}\dot{r}^2-r^2\dot{\theta}^2-r^2\sin^2\theta\dot{\phi}^2\right)\;,
\end{eqnarray}
where the dot denotes the derivative with respect to the $\tau$. Using the Euler-Lagrange equation, we obtain the equation of motion for the geodesic in the equatorial plane, $\theta={\pi}/{2}$,
\begin{eqnarray}\label{eoma}
\dot{r}^2+U\left(2\mathscr{L}+\frac{L^2}{r^2}\right)=E^2\;,
\end{eqnarray}
and
\begin{eqnarray}\label{eomb}
\dot{t}=\frac{E}{U}\;,\ \ \ \ \dot{\phi}=\frac{L}{r^2}\;,
\end{eqnarray}
where $E$ is the energy (inclusive of the rest energy) and $L$ is the angular momentum about the axis to the equatorial plane for the particle. By rescaling the affine parameter $\tau$, we can arrange that $2\mathscr{L}$ has the value of $+1$ for timelike geodesics. For null geodesics, $\mathscr{L}$ has the value of zero. By considering $r$ as the function of $\phi$ and letting $u=1/r$, we obtain the basic equation of the problem as in the analysis of Keplerian orbit in the Newtonian theory
\begin{eqnarray}\label{eom1}
&&\left(\frac{du}{d\phi}\right)^2=\frac{E^2-2\mathscr{L}}{L^2}+\frac{4\mathscr{L}Mu}{L^2}-\left(1+\frac{2\mathscr{L}Q^2}{L^2}\right)u^2
\nonumber\\&&+2Mu^3-Q^2u^4-\frac{4\mathscr{L}\alpha_2Q^4u^6}{5L^2}-\frac{2}{5}\alpha_2Q^4u^8\;.
\end{eqnarray}
This equation determines the geometry of the geodesics in the equatorial plane.
\section{The null geodesics}
\subsection{Radial null geodesics}
The equations governing the radial null-geodesics can be obtained by setting $L=0$ and $\mathscr{L}=0$ in Eq.~(\ref{eoma}) and Eq.~(\ref{eomb}), thus
\begin{eqnarray}
\dot{r}=\pm E\;,\ \ \  \dot{t}=\frac{E}{U}\;,\ \ \ \ \dot{\theta}=\dot{\phi}=0\;.
\end{eqnarray}
Therefore,
 \begin{eqnarray}\label{radial}
\frac{dr}{dt}=\pm U\;.
\end{eqnarray}
In order to solve this equation, we define the tortoise coordinate
\begin{eqnarray}
r_{\ast}&=&\int\frac{dr}{U}\;,
\end{eqnarray}
and rewrite $U$ as follows
\begin{eqnarray}
U&=&\left(1-\frac{r_1}{r}\right)\left(1-\frac{r_2}{r}\right)\left(1-\frac{r_3}{r}\right)\left(1-\frac{r_4}{r}\right)\nonumber\\&&\left(1-\frac{r_5}{r}\right)\left(1-\frac{r_6}{r}\right)\;.
\end{eqnarray}
Here $r_3,\ r_2, \ r_1$ denote the radii of three horizons. We assume $r_3>r_2>r_1>0$ and $r_4<0$. $r_5$ and $r_6$ are two conjugated complex numbers.
By using this form of $U$, we obtain
\begin{eqnarray}\label{tort}
r_{\ast}&=&\frac{r_1^6\ln|r-r_1|}{\left(r_1-r_2\right)\left(r_1-r_3\right)\left(r_1-r_4\right)\left(r_1-r_5\right)\left(r_1-r_6\right)}\nonumber\\
&&+\frac{r_2^6\ln|r-r_2|}{\left(r_2-r_1\right)\left(r_2-r_3\right)\left(r_2-r_4\right)\left(r_2-r_5\right)\left(r_2-r_6\right)}\nonumber\\
&&+\frac{r_3^6\ln|r-r_3|}{\left(r_3-r_1\right)\left(r_3-r_2\right)\left(r_3-r_4\right)\left(r_3-r_5\right)\left(r_3-r_6\right)}\nonumber\\
&&+\frac{r_4^6\ln|r-r_4|}{\left(r_4-r_1\right)\left(r_4-r_2\right)\left(r_4-r_3\right)\left(r_4-r_5\right)\left(r_4-r_6\right)}\nonumber\\
&&+\frac{r_5^6\ln|r-r_5|}{\left(r_5-r_1\right)\left(r_5-r_2\right)\left(r_5-r_3\right)\left(r_5-r_4\right)\left(r_5-r_6\right)}\nonumber\\
&&+\frac{r_6^6\ln|r-r_6|}{\left(r_6-r_1\right)\left(r_6-r_2\right)\left(r_6-r_3\right)\left(r_6-r_4\right)\left(r_6-r_5\right)}\nonumber\\
&&+r\;.
\end{eqnarray}
As defined,
\begin{eqnarray}
&&-\infty<r_\ast<+\infty\;,\ \ \ \ \textrm{for}\ \ \ \ r_3<r<+\infty\;,\nonumber\\
&&+\infty>r_\ast>-\infty\;,\ \ \ \ \textrm{for}\ \ \ \ r_2<r<r_3\;,\nonumber\\&&
-\infty<r_\ast<+\infty\;,\ \ \ \ \textrm{for}\ \ \ \ r_1<r<r_2\;,\nonumber\\&&
const>r_\ast>-\infty\;,\ \ \ \textrm{for}\ \ \ \ 0<r<r_1\;.
\end{eqnarray}
The solution to Eq.~(\ref{radial}) is
\begin{eqnarray}
t=\pm r_{\ast}+cosnt\;.
\end{eqnarray}
Therefore, for the in-going null-rays, the coordinate time $t$ increases from $-\infty$ to $+\infty$ as $r$ decreases from $+\infty$ to $r_{3}$,
decreases from $+\infty$ to $-\infty$ as $r$ further decreases from $r_{3}$ to $r_{2}$, increases from $-\infty$ to $+\infty$ as $r$ continually decreases from $r_2$ to $r_{1}$, and decreases finally from a finite value to $-\infty$ as $r$ decreases from $r_{1}$ to the black hole center. It should be noted that $dt/d\tau$ tends to $+\infty$ for $r\rightarrow r_{3}+0$, to $-\infty$ for $r\rightarrow r_{2}+0$ and to $+\infty$ for $r\rightarrow r_{1}+0$. Therefore, the radiation observed from infinity would appear infinitely red-shifted at the crossing of horizons $r_3$ and $r_1$, and infinitely blue-shifted at the crossing of horizon $r_2$.

\subsection{General null geodesics}

\begin{figure}[htbp]
	\centering
	\includegraphics[width=8cm,height=6cm]{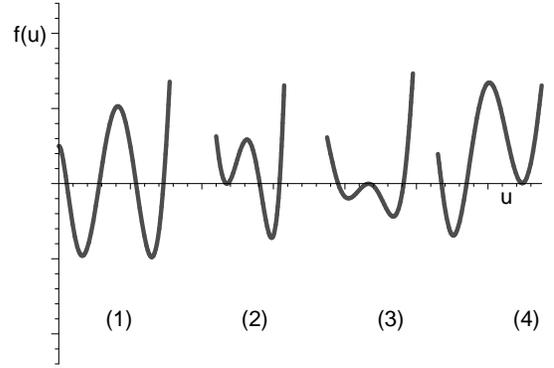}
	\centering
	\includegraphics[width=8cm,height=6cm]{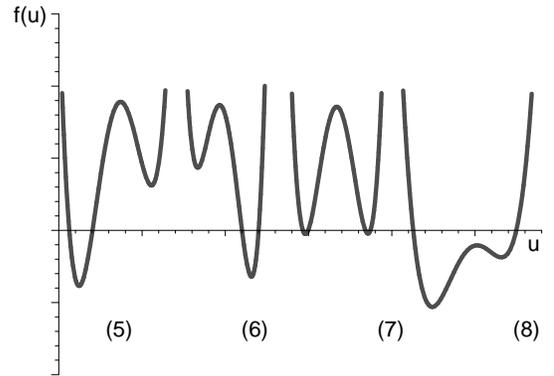}
    \centering
	\includegraphics[width=8cm,height=6cm]{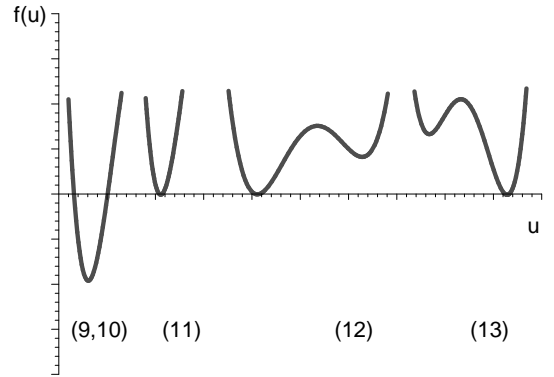}
	\caption{The disposition of the roots of the eight-degree equation $f(u)=0$ for $\alpha_2<0$.}
	\label{2210}
\end{figure}

Turning to the consideration of Eq.~(\ref{eom1}) with $\mathscr{L}=0$ for the general null geodesics. We have
\begin{eqnarray}\label{nulleom1}
\left(\frac{du}{d\phi}\right)^2&=&\frac{1}{D^2}-u^2
+2Mu^3-Q^2u^4\nonumber\\&&-\frac{2}{5}\alpha_2Q^4u^8\equiv f\left(u\right)\;,
\end{eqnarray}
where
\begin{eqnarray}
D\equiv\frac{L}{E}\;,
\end{eqnarray}
denotes the impact parameter. The property of orbits is determined by the number of roots of eight-degree equation, $f(u)=0$. In general, the equation has four distinguished and positive roots, $u_1,\ u_2, \ u_3, \ u_4$. But for some cases, double roots can arise. For example, $u_1$ and $u_2$ coincide. In this case, we denote $u_1=u_2$ with $u_{12}$. Taking into account of $\alpha_2<0$, we find we must consider $14$ cases as distinguished below:
\begin{eqnarray}
&&(1)\ \ \ \ u_1,\ \ \ u_2,\ \ \  u_3,\ \ \  u_4\;,\nonumber\\
&&(2)\ \ \ \ u_{12},\ \ \  u_3,\ \ \  u_4\;,\nonumber\\
&&(3)\ \ \ \ u_{1},\ \ \ u_{23},\ \ \  u_4\;,\nonumber\\
&&(4)\ \ \ \ u_{1},\ \ \ u_{2},\ \ \  u_{34}\;,\nonumber\\
&&(5)\ \ \ \ u_{1},\ \ \ u_{2}\;,\nonumber\\
&&(6)\ \ \ \ u_{3},\ \ \ u_{4}\;,\nonumber\\
&&(7)\ \ \ \ u_{12},\ \ \ u_{34}\;,\\
&&(8)\ \ \ \ u_{1},\ \ \ u_{4}\;,\nonumber\\
&&(9)\ \ \ \ u_{1},\ \ \ u_{234}\;,\nonumber\\
&&(10)\ \ \ \ u_{4},\ \ \ u_{123}\;,\nonumber\\
&&(11)\ \ \ \ u_{1234}\;,\nonumber\\
&&(12)\ \ \ \ u_{12}\;,\nonumber\\
&&(13)\ \ \ \ u_{34}\;,\nonumber\\
&&(14)\ \ \ \ \textrm{no real roots}\;.\nonumber
\end{eqnarray}
In Fig.~(\ref{2210}), we present the plots of $u$ vs $f(u)$ for these cases.
These different cases lead to the following conclusions.

\emph{Case} (1): The eight-degree equation, $f(u)=0$, allows four positive real roots with $0<u_1<u_2<u_3<u_4$. We must distinguish between orbits of the two kinds: the orbits of the first kind restricted to the interval, $0<u\leq u_1$ and $u_2\leq u\leq u_3$, and the second kind with $u\geq u_4$. The orbits in the interval, $0<u\leq u_1$, are the analogues of the hyperbolic orbits for massive particles in the Newtonian theory. The orbit in the interval, $u_2\leq u\leq u_3$, oscillates between two extreme values of $u=u_2$ and $u=u_3$ and they behave as the relativistic analogues of the Keplerian orbits for massive particles. The orbit of the second kind, starting at a certain aphelion distance, $u=u_4$, plunges into the singularity.

\emph{Case} (2): In this case, we have $0<u_1=u_2=u_{12}<u_3<u_4$ . The hyperbolic orbits and the oscillating orbits coalesce as they approach, asymptotically, a common circle from opposite sides by spiralling around it an infinite number of times. The orbit in the interval, $u\geq u_4$, starting at a certain aphelion distance, $u=u_4$, plunges into the singularity.

\emph{Case} (3): In this case, we have $0<u_1<u_2=u_3=u_{23}<u_4$. The orbit at $u=u_{23}$ is a stable circular orbit with zero eccentricity. The orbits at intervals, $0<u\leq u_1$ and $u\geq u_4$, are the same as the case (1).

\emph{Case} (4): In this case, we have $0<u_1<u_2<u_3=u_4=u_{34}$. The orbit of oscillation starts at a certain aphelion distance, $u_2$, and approaches the circle of radius $u_{4}^{-1}$, asymptotically, by spiraling around it an infinite number of times. The orbit of the second kind spirals away from the same circle and eventually plunges into the central singularity. The orbit at the  interval, $0<u\leq u_1$ is the same that in the case (1).

\emph{Case} (5): In this case, we have $0<u_1<u_2$ with $u_3,\ u_4$ a pair of complex-conjugate roots. The orbits belong to the hyperbolic and the plunging kind, respectively.

\emph{Case} (6): In this case, we have $0<u_3<u_4$ with $u_1,\ u_2$ a pair of complex-conjugate roots. The orbits belong to the hyperbolic and the plunging kind, respectively.

\emph{Case} (7): In this case, we have $0<u_1=u_2=u_{12}<u_3=u_4=u_{34}$. Then the hyperbolic orbits and the oscillating orbits coalesce as they approached, asymptotically, a common circle at radius, $u_{1}^{-1}$ ($=u_2^{-1}$), from opposite sides by spiralling round it an infinite number of times. On the other hand, the oscillating orbits and the plunging orbits also coalesce as they approached, asymptotically, a common circle at radius, $u_{3}^{-1}$ ($=u_4^{-1}$), from opposite sides by spiralling round it an infinite number of times.

\emph{Case} (8): In this case, we have $0<u_1<u_4$ with $u_2,\ u_3$ a pair of complex-conjugate roots. The orbits belong to the hyperbolic and the plunging kind, respectively.

\emph{Case} (9): In this case, we have $0<u_1<u_2=u_3=u_4=u_{234}$. The orbits include the hyperbolic orbits and the plunging orbits. A remarkable fact of the plunging orbits is that they have an unstable circular orbit with radius, $1/u_{234}$.

\emph{Case} (10): In this case, we have $0<u_1=u_2=u_3=u_{123}<u_4$. The orbits include the hyperbolic orbits and the plunging orbits. A remarkable fact of the hyperbolic orbits is that they have an unstable circular orbit with radius, $1/u_{123}$.

\emph{Case} (11): In this case, we have $0<u_1=u_2=u_3=u_4=u_{1234}$. Then the hyperbolic orbits and the plunging orbits coalesce as they approached, asymptotically, a common circle at radius, $u_{1234}^{-1}$, from opposite sides by spiralling round it an infinite number of times.

\emph{Case} (12): In this case, we have $0<u_1=u_2=u_{12}$ with $u_3,\ u_4$ a pair of complex-conjugate roots. The hyperbolic orbits and the plunging orbits coalesce as they approached, asymptotically, a common circle at radius, $u_{12}^{-1}$, from opposite sides by spiralling round it an infinite number of times.

\emph{Case} (13): In this case, we have $0<u_3=u_4=u_{34}$ with $u_1,\ u_2$ a pair of complex-conjugate roots. The hyperbolic orbits and the plunging orbits coalesce as they approached, asymptotically, a common circle at radius, $u_{34}^{-1}$, from opposite sides by spiralling round it an infinite number of times.

\emph{Case} (14): In this case, $f(u)=0$ has no real roots or $u_1,\ u_2$ a pair of complex-conjugate roots and $u_3,\ u_4$ a pair of complex-conjugate roots. The resulting orbits can be considered as belonging to imaginary eccentricities with the remarkable fact that they are unbound orbits. Like the bound orbits, these unbound orbits similarly fall into the central singularity, but are allowed to start from infinity rather than from finite aphelion distance.

\subsection{Stable circular orbits}
\emph{Case }(3) presents us a stable null circular orbit. We are interested in this orbit. The conditions for the occurrence of stable circular orbits are determined by
\begin{eqnarray}
f\left(u\right)=0\;, \ \ \textrm{and}\ \ \  f_{,u}=0\;,
\end{eqnarray}
where the comma denotes the derivative with respect to $u$. We note that these conditions are necessary but not sufficient. Observing the plot of case (3) in Fig.~(4), we assign the roots as $u_1=1,\ u_{23}=2,\ u_4=3$. Then we have
\begin{eqnarray}
f\left(u_1\right)&=&0\;, \ \ \ \ \ \ \ \  f\left({u_{23}}\right)=0\;,\nonumber\\
f\left(u_4\right)&=&0\;, \ \ \ \ \  f{,u\left(u_{23}\right)}=0\;.
\end{eqnarray}
Solving these equations, we obtain
\begin{eqnarray}
M&=&0.440\;, \ \ \ \ \  \ Q=0.466\;,\nonumber\\
D&=&1.724\;, \ \ \ \ \  \alpha_2=-0.020\;.
\end{eqnarray}
With above parameters, the equation of horizon, $U=0$, tells us there is only one event horizon with radius, $r_{EH}=0.462$. The radius of stable circular orbit is $1/u_{23}=0.5$. It is outside of the event horizon. \emph{{So we have a stable photon sphere outside the black hole.}}

\subsection{Null innermost stable circular orbit}
Different from Newtonian mechanics, the presence of an innermost stable circular orbit (ISCO) in General Relativity is a purely relativistic effect. The difference comes from the fact that the velocity of massive test particles cannot be equal or exceeding the speed of light. The ISCO represents the boundary between
test particles orbiting the black hole and test particles falling into the black hole. As a result, it marks the inner edge of the accretion disk in the accretion disk model of Shakura and Sunjaev \cite{sha:1973, abr:2013}. In accretion disk physics which can be compared to Event Horizon Telescopes observations, the model is frequently used as the starting point \cite{aki:2019}. Therefore, it is valuable to compute the ISCO for 3-horizon black holes.

The radius of null ISCO, $r_{_{SI}}=1/u_{_{SI}}$ is determined by
\begin{eqnarray}
f&=&0\;, \ \ \  f^{'}=0\;, \ \ \  f^{''}=0\;,\ \ \ f^{'''}=0\;,
\end{eqnarray}
where prime denotes the derivative with respect to $u$. Now we have four equations, but five unknown functions. So we are left with one freedom, for example, $u_{_{I}}$. Thus we obtain
\begin{eqnarray}
M&=&\frac{4}{5u_{_I}}\;,\label{78}\\
Q&=&\frac{\sqrt{3}}{2u_{_I}}\;,\label{79}\\
\alpha_2&=&-\frac{1}{9u_{_I}^2}\;,\label{80}\\
D&=&\frac{2\sqrt{2}}{u_{_I}}\;,
\end{eqnarray}
Now we make comparing of ISCO with Black hole event horizon. To this end, insert Eqs.~(\ref{78}), (\ref{79}) and (\ref{80}) into the horizon equation, we obtain
\begin{eqnarray}\label{ratio}
1-\frac{2M}{r_{_{EH}}}+\frac{Q^2}{r_{_{EH}}^2}+\frac{2}{5}\cdot\frac{\alpha_2^2Q^4}{r_{_{EH}}^6}=0\;.
\end{eqnarray}
Define the ratio for radii of null ISCO to event horizon, $\epsilon\equiv r_{_{I}}/r_{_{EH}}$, we find from above equation
\begin{eqnarray}
\epsilon\simeq1.5055\;,
\end{eqnarray}
which is larger than the Schwarzschild black hole (with $\epsilon=3/2$). \emph{The radius of null ISCO determines the boundary of
the black hole shadow. Thus we conclude that the shadow of a 3-horizon black hole is larger than that of a Schwarzschild black hole although they have the same size of event horizon.}

\section{Timelike geodesics}
\subsection{The radial geodesics}
In this subsection, we shall consider the radial timelike geodesics with zero angular momentum. The equations governing these geodesics
are
\begin{eqnarray}
\left(\frac{dr}{d\tau}\right)^2=E^2-U\;, \ \ \ \textrm{and}\ \ \ \ \   \frac{dt}{d\tau}=\frac{E}{U}\;.
\end{eqnarray}
Since $U>0$ in the interval $r_1<r<r_2$, it is obvious that $E^2-U$ will vanish for some finite value of $r_1<r<r_2$ provided that $E^2$ is smaller than the maximum of $U$ for the interval $r_1<r<r_2$. We conclude that the trajectory will have a turning point in the interval $r_1<r<r_2$ in this case. On the other hand, if $E^2$ is larger than the maximum of $U$ for the interval $r_1<r<r_2$, the radial timelike trajectory would reach the singularity.
\subsection{The general timelike geodesics}
\begin{figure}[htbp]
		\includegraphics[width=8cm,height=6cm]{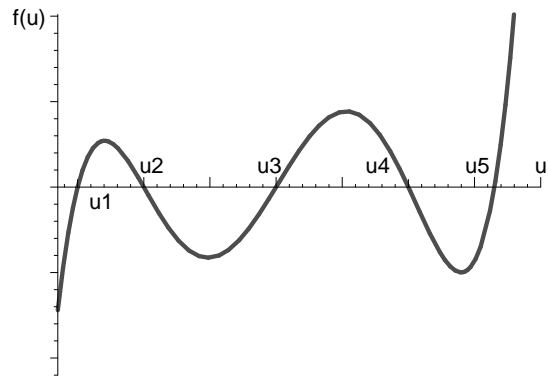}
		\caption{The disposition of the five roots of the eight-degree equation $f(u)=0$ for general timelike geodesics when $\alpha_2<0$.}
	\label{12345}
\end{figure}

Turning to the consideration of Eq.~(\ref{eom1}) with $2\mathscr{L}=1$ for the general timelike geodesics. We have
\begin{eqnarray}
&&\left(\frac{du}{d\phi}\right)^2=\frac{E^2-1}{L^2}+\frac{2Mu}{L^2}-\left(1+\frac{Q^2}{L^2}\right)u^2
+2Mu^3\nonumber\\&&-Q^2u^4-\frac{2\alpha_2Q^4u^6}{5L^2}-\frac{2}{5}\alpha_2Q^4u^8\equiv f\left(u\right)\;.
\end{eqnarray}
When $E^2>1$ and $\alpha_2<0$, the eight-degree equation, $f(u)=0$, allows at most four positive real roots with $0<u_1<u_2<u_3<u_4$. The cases we must consider are exactly the eight cases distinguished in Fig.~(\ref{2210}). Therefore, we shall not consider them any more.
However, when $E^2<1$ and $\alpha_2<0$, $f(u)=0$ can have at most five positive roots with $0<u_1<u_2<u_3<u_4<u_5$ as shown in Fig.~(\ref{12345}).

We find that we must consider $25$ cases distinguished below:
\begin{eqnarray}
&&(1)\ \ \ \ u_1,\ \ \ u_2,\ \ \  u_3,\ \ \  u_4,\ \ \ u_5\;,\nonumber\\
&&(2)\ \ \ \ u_{12},\ \ \  u_3,\ \ \  u_4,\ \ \ u_5\;,\nonumber\\
&&(3)\ \ \ \ u_1,\ \ \ u_{23},\ \ \  u_4,\ \ \ u_5\;,\nonumber\\
&&(4)\ \ \ \ u_1,\ \ \ u_2,\ \ \  u_{34},\ \ \ u_5\;,\nonumber\\
&&(5)\ \ \ \ u_1,\ \ \ u_2,\ \ \  u_3,\ \ \  u_{45}\;,\nonumber\\
&&(6)\ \ \ \ u_{123},\ \ \  u_4,\ \ \ u_5\;,\nonumber\\
&&(7)\ \ \ \ u_{12},\ \ \  u_{34},\ \ \ u_5\;,\nonumber\\
&&(8)\ \ \ \ u_{12},\ \ \  u_3,\ \ \  u_{45}\;,\nonumber\\
&&(9)\ \ \ \ u_1,\ \ \ u_{234},\ \ \ u_5\;,\nonumber\\
&&(10)\ \ \ \ u_1,\ \ \ u_{23},\ \ \  u_{45}\;,\nonumber\\
&&(11)\ \ \ \ u_1,\ \ \ u_2,\ \ \  u_{345}\;,\nonumber\\
&&(12)\ \ \ \ u_1,\ \ \ u_2,\ \ \  u_{5}\;,\nonumber\\
&&(13)\ \ \ \ u_3,\ \ \ u_4,\ \ \  u_{5}\;,\\
&&(14)\ \ \ \ u_1,\ \ \ u_2,\ \ \  u_{3}\;,\nonumber\\
&&(15)\ \ \ \ u_1,\ \ \ u_4,\ \ \  u_{5}\;,\nonumber\\
&&(16)\ \ \ \ u_{1234},\ \ \ u_5\;,\nonumber\\
&&(17)\ \ \ \ u_{123},\ \ \  u_{45}\;,\nonumber\\
&&(18)\ \ \ \ u_{12},\ \ \  u_{345}\;,\nonumber\\
&&(19)\ \ \ \ u_1,\ \ \ u_{2345}\;,\nonumber\\
&&(20)\ \ \ \ u_{12},\ \ \ u_{5}\;,\nonumber\\
&&(21)\ \ \ \ u_{34},\ \ \ u_{5}\;,\nonumber\\
&&(22)\ \ \ \ u_{23},\ \ \ u_{1}\;,\nonumber\\
&&(23)\ \ \ \ u_{45},\ \ \ u_{1}\;,\nonumber\\
&&(24)\ \ \ \ u_{123}\ \ \textrm{or}\ \ u_{345}\ \ \textrm{or}\ \ u_{12345}\;,\nonumber\\
&&(25)\ \ \ \ u_{1}\ \ \textrm{or} \ \ u_5\;.\nonumber
\end{eqnarray}

\emph{Case} (1): The eight-degree equation, $f(u)=0$, allows five positive real roots with $0<u_1<u_2<u_3<u_4<u_5$. We must distinguish between orbits of the two kinds: the orbits of the first kind restricted to the interval $u_1\leq u\leq u_2$ and $u_3\leq u\leq u_4$, and the second kind with $u\geq u_5$. The orbit in the interval, $u_1\leq u\leq u_2$ and $u_3\leq u\leq u_4$, oscillates between two extreme values of $u=u_1$, $u=u_2$ and $u=u_3$, $u=u_4$, respectively. They are the relativistic analogues of the Keplerian orbits for massive particles. The orbit of the second kind, starting at a certain aphelion distance, $u=u_5$, plunges into the singularity.

\emph{Case} (2): In this case, $u_1$ and $u_2$ coincide and there exists stable circular orbit with radius $1/u_{12}$ except for the oscillating and plunging orbits.

\emph{Case} (3): In this case, $u_2$ and $u_3$ coincide and there exists unstable circular orbit with radius $1/u_{23}$. The two oscillating orbit coalesce as they approach, asymptotically, the common circle with radius $1/u_{23}$, from opposite sides by spiralling around it an infinite number of times.

\emph{Case} (4): In this case, $u_3$ and $u_4$ coincide and there exists stable circular orbit with radius $1/u_{34}$ except for the oscillating and plunging orbits.

\emph{Case} (5): In this case, $u_4$ and $u_5$ coincide and there exists unstable circular orbit with radius $1/u_{45}$. The oscillating orbit and the plunging orbits coalesce as they approach, asymptotically, the common circle with radius $1/u_{45}$, from opposite sides by spiralling around it an infinite number of times.

\emph{Case} (6): In this case, $u_1,\ u_2$ and $u_3$ coincide and there exists unstable circular orbit with radius $1/u_{123}$ except for the oscillating and plunging orbits.

\emph{Case} (7): In this case, $u_1$ and $u_2$ coincide while $u_3$ and $u_4$ coincide. There exist two stable circular orbits with radius $1/u_{12}$ and $1/u_{34}$, respectively, except for the plunging orbits.

\emph{Case} (8): In this case, $u_1$ and $u_2$ coincide while $u_4$ and $u_5$ coincide. There exists a stable circular orbit with radius $1/u_{12}$ and am unstable circular orbit with radius $1/u_{45}$, respectively. The oscillating orbit in the interval, $u_{3}\leq u\leq u_{45}$ and the plunging orbits in the interval, $u\geq u_{45}$ coalesce as they approach, asymptotically, the common circle with radius $1/u_{45}$, from opposite sides by spiralling around it an infinite number of times.

\emph{Case} (9): In this case, $u_2, \ u_3$ and $u_4$ coincide and there exists unstable circular orbit with radius $1/u_{234}$ except for the oscillating and plunging orbits.

\emph{Case} (10): In this case, $u_2$ and $u_3$ coincide while $u_4$ and $u_5$ coincide. There exist two unstable circular orbits with radius $1/u_{23}$ and $1/u_{45}$, respectively, except for the  oscillating and the plunging orbits.

\emph{Case} (11): In this case, $u_3,\ u_4$ and $u_5$ coincide and there exists an unstable circular orbit with radius $1/u_{345}$ except for the oscillating orbit in the interval, $u_{1}\leq u\leq u_{2}$ and the plunging orbits in the interval, $u\geq u_{345}$.

\emph{Case} (12): In this case, there are three real roots, $u_{1}$, $u_{2}$ and $u_{5}$. $u_3$ and $u_4$ are a pair of complex-conjugate roots. So there exist the oscillating orbit in the interval, $u_{1}\leq u\leq u_{2}$ and the plunging orbits in the interval, $u\geq u_{5}$.

\emph{Case} (13): In this case, there are three real roots, $u_{3}$, $u_{4}$ and $u_{5}$. $u_1$ and $u_2$ are a pair of complex-conjugate roots. Similar to case (12), there exist the oscillating orbit in the interval, $u_{3}\leq u\leq u_{4}$ and the plunging orbits in the interval, $u\geq u_{5}$.

\emph{Case} (14): In this case, there are three real roots, $u_{1}$, $u_{2}$ and $u_{3}$. $u_4$ and $u_5$ are a pair of complex-conjugate roots. Similar to case (12), there exist the oscillating orbit in the interval, $u_{1}\leq u\leq u_{2}$ and the plunging orbits in the interval, $u\geq u_{3}$.

\emph{Case} (15): In this case, there are three real roots, $u_{1}$, $u_{4}$ and $u_{5}$. $u_2$ and $u_3$ are a pair of complex-conjugate roots. Similar to case (12), there exist the oscillating orbit in the interval, $u_{1}\leq u\leq u_{4}$ and the plunging orbits in the interval, $u\geq u_{5}$.

\emph{Case} (16): In this case, $u_1,
\ u_2,\ u_3,\ u_4$ coincide and there exists a stable circular orbit with radius $1/u_{1234}$ except for the plunging orbits in the interval, $u\geq u_{5}$.

\emph{Case} (17): In this case, $u_1,\ u_2$ and $u_3$ coincide while $u_4$ and $u_5$ coincide. There exist two  unstable circular orbits with radius $1/u_{123}$ and $1/u_{45}$, respectively, except for the  oscillating and the plunging orbits.

\emph{Case} (18): In this case, $u_1$ and $u_2$ coincide while $u_3, \ u_4$ and $u_5$ coincide. There exists a stable circular orbit with radius $1/u_{12}$ and am unstable circular orbit with radius $1/u_{345}$, respectively, except for the plunging orbits.

\emph{Case} (19): In this case, $u_2,\ u_3,\ u_4,\ u_5$ coincide. There exists an unstable circular orbit with radius $1/u_{2345}$  except for the  oscillating and the plunging orbits.

\emph{Case} (20): In this case, there are two real roots with double toots $u_{12}$ and $u_{5}$. $u_3$ and $u_4$ are a pair of complex-conjugate roots. There is a stable circular orbit with radius $1/u_{12}$ and the plunging orbits in the interval, $u\geq u_{5}$.

\emph{Case} (21): In this case, there are two real roots with double toots $u_{34}$ and $u_{5}$. $u_1$ and $u_2$ are a pair of complex-conjugate roots. Similar to case (20), there is a stable circular orbit with radius $1/u_{34}$ and the plunging orbits in the interval, $u\geq u_{5}$.

\emph{Case} (22):  In this case, there are two real roots with double toots $u_{23}$ and $u_{1}$. $u_4$ and $u_5$ are a pair of complex-conjugate roots. The oscillating orbits stars at a certain aphelion distance, $1/u_{1}$, and approaches the circle of radius $1/u_{23}$,asymptotically, by spiralling around it an infinite number of times. The orbit of the plunging is, in some sense, a continuation of the oscillating orbits in that it spirals away from the circle and then plunge eventually into the central singularity.

\emph{Case} (23): In this case, there are two real roots with double toots $u_{45}$ and $u_{1}$. $u_2$ and $u_3$ are a pair of complex-conjugate roots. The properties of orbits are similar to the case (22).

\emph{Case} (24): In this case, there is only one real root, $u_{123}$ or $u_{345}$ or $u_{12345}$. There exists an unstable circular orbit with radius $1/u_{123}$,  $1/u_{345}$ or $1/u_{12345}$ except for the  plunging orbits.

\emph{Case} (25): In this case, there is only one real root, $u_1$ or $u_5$ (the other four roots are complex-conjugate roots). We have only one class of orbits. Namely, all the orbits plunge into the singularity after starting from certain aphelion distances.

\subsection{Stable timelike circular orbits}
\emph{Case }(7) presents us two stable timelike circular orbits with radius $1/u_{12}$ and $1/u_{34}$, respectively. We consider the orbits in this subsection. The conditions for the occurrence of stable circular orbits are determined by
\begin{eqnarray}
f\left(u\right)=0\;, \ \ \textrm{and}\ \ \  f_{,u}=0\;,
\end{eqnarray}
where the comma denotes the derivative with respect to $u$. We assign the roots as $u_{12}=1,\ u_{34}=2,\ u_5=3$. Then we have
\begin{eqnarray}
f\left(u_{12}\right)&=&0\;, \ \ \textrm{}\ \ \ \ \ \ f\left({u_{34}}\right)=0\;,\nonumber\\
f\left(u_{5}\right)&=&0\;, \ \ \textrm{}\ \ \  f{,u\left(u_{12}\right)}=0\;.\nonumber\\
f{,u\left(u_{34}\right)}&=&0\;.
\end{eqnarray}
Solving these equations, we obtain
\begin{eqnarray}
M&=&0.299\;, \ \ \ \ \  Q=0.335\;,\nonumber\\
E&=&0.899\;, \ \ \ \ \  L=0.756\;,\nonumber\\
\alpha_2&=&-0.023\;.
\end{eqnarray}
With above parameters, the equation of horizon, $U=0$, tells us there is only one event horizon with radius, $r_{EH}=0.264$. The radius of two stable circular orbits are $1/u_{12}=1$ and $1/u_{34}=0.5$, respectively. They are all larger than the event horizon. \emph{So we have two stable matter spheres outside the black hole.}

\subsection{Timelike innermost stable circular orbit}
The radius of timelike ISCO, $r_{_{SI}}=1/u_{_{SI}}$ is determined by
\begin{eqnarray}
f&=&0\;, \ \ \  f^{'}=0\;, \ \ \  f^{''}=0\;,\nonumber\\
f^{'''}&=&0\;, \ \ \  f^{(4)}=0\;,
\end{eqnarray}
where prime denotes the derivative with respect to $u$. We have five equations, but six unknown functions. We choose $u_{_{I}}$ as the remaining freedom. Thus we obtain

\begin{eqnarray}
M&=&\frac{4}{415}\cdot\frac{2357+367\sqrt{41}}{\left(47+7\sqrt{41}\right)u_{_{I}}}\;,\label{83}\\
Q&=&\frac{1}{166}\cdot\frac{7636+166\sqrt{41}}{u_{_{I}}}\;,\label{84}\\
\alpha_2&=&-\frac{83}{2}\cdot\frac{5+\sqrt{41}}{\left(46+\sqrt{41}\right)\left(47+7\sqrt{41}\right)u_{_{I}}^2}\;,\label{85}\\
L&=&\frac{1}{4}\cdot\frac{\sqrt{10+2\sqrt{41}}}{u_{_{I}}}\;,\\
E&=&\frac{1}{664}\cdot\sqrt{389602-6808\sqrt{41}}\;.
\end{eqnarray}
Notice that the energy of the particles is independent on the inverse radius $u_{_{I}}$ of ISCO. In order to compare the ISCO with black hole event horizon, we insert Eqs.~(\ref{83}),(\ref{84}) and (\ref{85}) into the horizon equation and eventually find
\begin{eqnarray}
\epsilon\simeq2.0678\;,
\end{eqnarray}
which is smaller than the Schwarzschild black hole (with $\epsilon=3$). \emph{This means the inner edge of accretion disk for the 3-horizon black holes is pushed more closer to the event horizon compared with the Schwarzschild black hole.}

\section{The motion of charged particles}
The motion of a test particle with net charge is determined by the Lagrangian
\begin{eqnarray}
\mathscr{L}
=\frac{1}{2}\left(U\dot{t}^{2}-\frac{1}{U}\dot{r}^2-r^2\dot{\theta}^2-r^2\sin^2\theta\dot{\phi}^2\right)+{q\Phi}\dot{t}\;,
\end{eqnarray}
where $q$ denotes the charge per unit mass of the test particle. The equations of motion in the equatorial plane following from this Lagrangian are

\begin{eqnarray}
U\dot{t}+{q\Phi}=E\;,\ \  \ r^2\dot{\phi}=L\;,
\end{eqnarray}
and
\begin{eqnarray}\label{eom1a}
\dot{r}^2+U\left(1+\frac{L^2}{r^2}\right)=\left(E-q\Phi\right)^2\;,
\end{eqnarray}
and in place of Eq.~(\ref{eom1}) we obtain

\begin{eqnarray}\label{eom2}
&&\left(\frac{du}{d\phi}\right)^2=\frac{E^2-1}{L^2}+\frac{2\left(M-qEQ\right)u}{L^2}\nonumber\\&&-\left[1+\frac{Q^2\left(1-q^2\right)}{L^2}\right]u^2+2Mu^3
-Q^2u^4\nonumber\\&&-\frac{8qE\alpha_2Q^3u^5}{5L^2}-\frac{2\alpha_2Q^4\left(1-4q^2\right)u^6}{5L^2}\nonumber\\&&-\frac{2}{5}\alpha_2Q^4u^8+\frac{16q^2\alpha_2^2Q^6u^{10}}{25L^2}
\equiv f\left(u\right)\;.
\end{eqnarray}
We can do orbital analysis as in previous sections. But we will not do it here. One novel feature of Eq.~(\ref{eom1a}) is that when the test particle has a turning point at it arrives at the event horizon, its energy will be

\begin{eqnarray}
E=q\Phi\mid_{r=r_{EH}}=\frac{qQ}{r_{EH}}+\frac{4\alpha_2 qQ^3}{5}\cdot\frac{1}{r_{EH}^5}\;.
\end{eqnarray}
In the absence of $\alpha_2$, the energy is negative if and only if the charges $q$ and $Q$ have different signs. However, in the presence of $\alpha_2$,
the energy can be negative even if the charges $q$ and $Q$ have the same signs provided that $\alpha_2$ is sufficiently negative. The negative of energy $E$  leads to the Penrose process that one can extract energy from the black hole by using charged particles.

\section{static response and Love numbers}
In this section, we shall study the static response of the 3-horizon black holes to external scalar field. The quantities of response are
intrinsic and contribute to the form of gravitational waves. Therefore, they can be in principle discovered \cite{car:2017} in the observation of gravitational waves.
The response of an object to a long-wavelength tidal field is encoded in the so-called Love
numbers \cite{lov:1909}, which describe the deformability or rigidity of the object. The Love numbers of Schwarzschild black hole are
exactly vanishing in four dimensional spacetime \cite{dam:2009,{bin:2009},{fang:2005},kol:2012,{cha:2013},gur:2015,hui:2021}. This reveals the Schwarzschild black hole is totally rigid. However, it is not case when one consider higher
dimensions \cite{kol:2012,car:2019}, anti-de Sitter asymptotical \cite{em:2017}, the presence of higher-curvature terms \cite{car:2017}, or
many alternative theories of gravity \cite{car:2017a}. It is found that the Love numbers of those cases are non-vanishing. Here we compute the Love numbers of 3-horizon black hole caused by scalar tidal field. The 3-horizon black hole can be thought as the extension of RN black hole and to our
knowledge, one did not compute the Love numbers for RN black holes. So in the next, we shall begin from the studying of RN black holes.

\subsection{The Love numbers of extremal Reissner-Nordstrom black holes}
The extremal RN metric is
\begin{eqnarray}
ds^2=-\frac{\Delta}{r^2}dt^2+\frac{r^2}{\Delta}dr^2+r^2\left(d\theta^2+\sin^2\theta d\phi^2\right)\;,
\end{eqnarray}
where $\Delta=r^2-2Mr+M^2$ with $M$ the mass of the black hole. This spacetime has only one event horizon with the radius $r_{EH}=M$.
A static, massless scalar field in this background
satisfies the equation of motion \cite{hui:2021,hui:2021a}
\begin{eqnarray}
\partial_r\left(\Delta\partial_r{\phi_{l}}\right)-l\left(l+1\right)\phi_l=0\;,
\end{eqnarray}
where $\phi_l$ is the radial component of the scalar field $\Phi$ which has been decomposed as $\Phi=\phi_l(r)Y_{lm}(\theta,\ \phi)$. $Y_{lm}(\theta,\ \phi)$ is the spherical harmonic function. $l$ is the angular quantum number. Make the change of variables, $r\rightarrow z$ as follows

\begin{eqnarray}
z=-\int\frac{1}{\Delta}dr=\frac{1}{r-M}\;,
\end{eqnarray}
namely,
\begin{eqnarray}
r=M+\frac{1}{z}\;.
\end{eqnarray}
The region of $r\in[r_{_{EH}},\ +\infty)$ is mapped into $z\in[+\infty,\ 0)$. Then the equation of motion becomes

\begin{eqnarray}
\frac{d^2\phi_l}{dz^2}-\frac{l\left(l+1\right)}{z^2}\phi_{l}=0\;,
\end{eqnarray}

Solving this equation, we obtain

\begin{eqnarray}
\phi_{l}=c_{1}z^{l+1}+c_{2}z^{-l}\;,
\end{eqnarray}
 or
 \begin{eqnarray}
\phi_{l}=c_{1}\left(r-M\right)^{-l-1}+c_{2}\left(r-M\right)^{l}\;,
\end{eqnarray}
where $c_1$ and $c_2$ are integration constants. Now we require two boundary conditions to
specify completely the solution. The first boundary condition is that the scalar field is finite at the event horizon. The second boundary condition is to fix the normalization of the growing mode solution at spatial infinity. Then one can read off the induced sub-leading fall-off at spatial infinity, which plays the role of the linear response to the external field. The Love number, describing the response of the black hole to external perturbations, is defined as the ratio between the coefficients of the decaying and growing modes at infinity.

The first boundary condition reveals $c_1=0$. The second boundary condition tells us we have $\phi_{l}=c_{2}r^{l}$ in spatial infinity. We can understand this point by imagining that we are applying a scalar field, in which the black hole is immersed, that scales like $r^l$ as $r\rightarrow +\infty$ with angular structure given by $l=1,2,3,\cdot\cdot\cdot$ harmonics.

Therefore, the long-wavelength external scalar field one can apply is
 \begin{eqnarray}
\phi_{l}=c_{2}\left(r-M\right)^{l}\;.
\end{eqnarray}
This solution is a pure growing mode and the sub-leading induced fall-off mode does not exist. This means the Love numbers of extreme RN black holes
are vanishing.

\subsection{The Love numbers of non-extremal Reissner-Nordstrom black holes}
In this case, we have $\Delta=(r-r_+)(r-r_1)$ with $r_{\pm}=M\pm\sqrt{M^2-Q^2}$.
Make the change of variables, $r\rightarrow z$ as follows

\begin{eqnarray}
z=-\int\frac{1}{\Delta}dr=\frac{1}{r_+-r_{-}}\ln{\mid\frac{r-r_{-}}{r-r_{+}}\mid}\;,
\end{eqnarray}
namely,
\begin{eqnarray}
r=\frac{r_+-r_{-}e^{\left(r_+-r_{-}\right)z}}{e^{\left(r_+-r_{-}\right)z}-1}\;.
\end{eqnarray}
The region of $r\in[r_{+},\ +\infty)$ is mapped into $z\in[+\infty,\ 0)$. The equation of motion becomes

\begin{eqnarray}
\frac{d^2\phi_L}{dz^2}-\frac{L\left(L+1\right)\left(r_+-r_{-}\right)^2}{4\sinh^2\frac{\left(r_+-r_{-}\right)z}{2}}\phi_{L}=0\;.
\end{eqnarray}

Solving this equation, we obtain
\begin{eqnarray}
\phi_{L}&=&c_{1}\mathfrak{L}+c_2\mathfrak{L}\int\frac{dz}{\mathfrak{L}^2}\;,
\end{eqnarray}
where $\mathfrak{L}$ is defined by
 \begin{eqnarray}
\mathfrak{L}&=&{\textrm{LegendreP}}\left(L,\  \coth{\frac{\left(r_+-r_{-}\right)z}{2}}\right)\;.
\end{eqnarray}
$c_1$ and $c_2$ are integration constants. We have $z=+\infty$ on the event horizon $r=r_+$. When $z\rightarrow +\infty$, we have
\begin{eqnarray}
\mathfrak{L}&=&1\;.
\end{eqnarray}
The first boundary tells us $c_2$ term must be dropped. Therefore, the
long-wavelength external scalar field can be applied is
 \begin{eqnarray}
\phi_{L}=c_{1}\mathfrak{L}\;.
\end{eqnarray}
It is a growing mode with respect to $r$ and the decaying mode does not exist. Thus the Love number of non-extremal RN black hole is vanishing.

\subsection{Love number of 3-horizon black hole}
In this case, we have $\Delta=(r-r_1)(r-r_2)(r-r_3)(r-r_4)(r-r_5)(r-r_6)/r^4$.
Here $r_3,\ r_2, \ r_1$ denote the radii of three horizons. We assume $r_3>r_2>r_1>0$ and $r_4<0$. $r_5$ and $r_6$ are two conjugated complex numbers.

Make the change of variables, $r\rightarrow z$ as follows
\begin{eqnarray}\label{tort}
z&=&-\int\frac{1}{\Delta}dr\nonumber\\&=&-\frac{r_1^4\ln|r-r_1|}{\left(r_1-r_2\right)\left(r_1-r_3\right)\left(r_1-r_4\right)\left(r_1-r_5\right)\left(r_1-r_6\right)}\nonumber\\
&&-\frac{r_2^4\ln|r-r_2|}{\left(r_2-r_1\right)\left(r_2-r_3\right)\left(r_2-r_4\right)\left(r_2-r_5\right)\left(r_2-r_6\right)}\nonumber\\
&&-\frac{r_3^4\ln|r-r_3|}{\left(r_3-r_1\right)\left(r_3-r_2\right)\left(r_3-r_4\right)\left(r_3-r_5\right)\left(r_3-r_6\right)}\nonumber\\
&&-\frac{r_4^4\ln|r-r_4|}{\left(r_4-r_1\right)\left(r_4-r_2\right)\left(r_4-r_3\right)\left(r_4-r_5\right)\left(r_4-r_6\right)}\nonumber\\
&&-\frac{r_5^4\ln|r-r_5|}{\left(r_5-r_1\right)\left(r_5-r_2\right)\left(r_5-r_3\right)\left(r_5-r_4\right)\left(r_5-r_6\right)}\nonumber\\
&&-\frac{r_6^4\ln|r-r_6|}{\left(r_6-r_1\right)\left(r_6-r_2\right)\left(r_6-r_3\right)\left(r_6-r_4\right)\left(r_6-r_5\right)}\;.\nonumber\\
\end{eqnarray}
Then the region of $r\in[r_{3},\ +\infty)$ is mapped into $z\in[+\infty, \ 0)$. The equation of motion becomes

\begin{eqnarray}\label{eom-3}
\frac{d^2\phi_L}{dz^2}-L\left(L+1\right)\Delta\left(z\right)\phi_{L}=0\;.
\end{eqnarray}
Here $\Delta$ is understood as the function of $z$. Now we look for the fall-off solution at $r\rightarrow +\infty$, or $z\rightarrow 0$. When $r\rightarrow +\infty$, we have $\Delta=1/z^2$ and $r=1/z$. The equation of motion, Eq.~(\ref{eom-3}), reduces to exactly that for the extremal RN black hole. The corresponding Love number is zero. On the other hand, when  $r\rightarrow r_3$, or $z\rightarrow +\infty$, we have $\Delta=c_3\left(r-r_5\right)$ with $c_3$ a positive constant and $r=r_3+e^{-c_3 z}$. The resulting equation of motion is
\begin{eqnarray}\label{eom-3a}
\frac{d^2\phi_L}{dz^2}-L\left(L+1\right)c_3 e^{-c_3 z}\phi_{L}=0\;.
\end{eqnarray}
The solution is

 \begin{eqnarray}
\phi_{L}&=&c_{1}\textrm{BesselI}\left(0,\ \frac{2\sqrt{L\left(L+1\right)}}{\sqrt{c_3 e^{c_3 z}}}\right)\nonumber\\&&
+c_{2}\textrm{BesselK}\left(0,\ -\frac{2\sqrt{L\left(L+1\right)}}{\sqrt{c_3 e^{c_3 z}}}\right)\;.
\end{eqnarray}

The $c_2$ term is divergent on the horizon. Therefore it should be dropped. We eventually find that
\begin{eqnarray}
\phi_{L}&=&c_{1}\textrm{BesselI}\left(0,\ \frac{2\sqrt{L\left(L+1\right)}}{\sqrt{c_3 e^{c_3 z}}}\right)\;.
\end{eqnarray}
It is a growing mode with  $\phi_{L}=c_1$ when ${r=r_3}$. There is no the decaying fall-off mode. In other words, the Love number of 3-horizon black hole is also vanishing.
\section{Thermodynamics}
Finally, we make an investigation of the thermodynamics for the 3-horizon black hole. Concretely, we shall derive the Smarr formula and the first law of thermodynamics. It is worth noting that many studies in NLE in this respect have been carried out, for example in \cite{hos:2015,gul:2018,gul:2021,bre:2005,bal:2017,wan:2019}. Hawking showed that for the outermost, event horizon in an asymptotically flat spacetime, the
temperature of black hole is
\begin{equation}
T_{EH}=\frac{\kappa_{EH}}{2\pi}\;,
\end{equation}
where the surface gravity $\kappa_{EH}$ is defined by evaluating
\begin{equation}
l^{\mu}\nabla_{\mu}l^{\nu}=\kappa l^{\nu}\;,
\end{equation}
on the event horizon. Here $l^{\mu}$ is the future-directed null generator of the event horizon, which coincides
with a Killing vector $K^{\mu}$ on the horizon. The metric for the 3-horizon black holes is static. So if the Killing vector is adopted as $K^{\mu}=\partial/\partial t$, then we have
\begin{equation}
\kappa_{EH}=\frac{1}{2}\cdot\frac{dU}{dr}\mid_{r=r_{EH}}\;.
\end{equation}

As a result, the black hole temperature takes the form
\begin{eqnarray}
T_{EH}&=&\frac{1}{2\pi}\left(\frac{M}{r_{EH}^2}-\frac{Q^2}{r_{EH}^3}-\frac{6}{5}\frac{\alpha_2Q^4}{r_{EH}^7}\right)\;.
\end{eqnarray}
The radius $r_{EH}$ of event horizon is determined by
 \begin{eqnarray}\label{eq:t1}
U\mid_{r=r_{EH}}=0\;.
\end{eqnarray}
Thus the temperature is eventually determined by the mass $M$, charge $Q$ and coupling constant $\alpha_2$.

The entropy of black holes generally satisfies the area law which states that the entropy is a quarter of the area of black hole event horizon \cite{beck:1973,haw:1974,gib:1977}.
Therefore we have the entropy of the black hole
 \begin{equation}
S_{EH}=\pi r_{EH}^2\;.
\end{equation}

The electrostatic potential on the event horizon is
\begin{equation}\label{eq:p1}
\Phi_{EH}=\Phi\mid_{r=r_{EH}}\;.
\end{equation}

We find if we define the pressure $P$ and the thermodynamic volume $\mathfrak{V}$ as follows, respectively,
\begin{equation}
P\equiv\frac{1}{\alpha_2}\;,
\end{equation}
\begin{equation}\label{eq:v1}
\mathfrak{V}\equiv\left(\frac{\partial M}{\partial P}\right)_{S,Q}=-\frac{\alpha_2^2Q^4}{5r_{EH}^5}\;.
\end{equation}
Then the Smarr formula
\begin{equation}
M=2T_{EH}S_{EH}-2\mathfrak{V}P+Q\Phi_{EH}\;,
\end{equation}
is satisfied. We could make an examination on whether the thermal quantities fulfill the requirement of the first law of thermodynamics.
To this end, we treat the mass $M$, the entropy $S$ and the pressure $P$ as the function of $r_{EH},\ Q,\ P$. Then we have
\begin{equation}
dM=M_{,r_{EH}}dr_{EH}+M_{,Q}dQ+M_{,P}dP\;,
\end{equation}
\begin{equation}
dS=S_{,r_{EH}}dr_{EH}\;.
\end{equation}
After computation, we find the first law of thermodynamics
\begin{equation}
dM=T_{EH}dS_{EH}+\Phi_{EH}dQ+\mathfrak{V}dP\;,
\end{equation}
is indeed satisfied.

\section{{Conclusion and discussion}}

In summary, starting from the NLE Lagrangian with infinite series of Maxwell invariant and using the method of infinite series, we find the black hole solutions with many horizons. To be specific, we present the solutions for 3-horizon and 4-horizon black holes. In particular, the explicit and analytic expression for NLE Lagrangian of the 3-horizon black hole is obtained. To our knowledge, one did not yet get 3-horizon black holes with an analytic NLE Lagrangian. On the other hand, our lagrangians have intrinsics advantage over previous studies \cite{noj:2017,gao:2018}.  They are not dependent on the mass $M$, charge $Q$, or the ratio between charge to mass $Q/M$. There are only related to the coupling constants, $\alpha_i$. We find that there are three physical parameters, the mass $M$, charge $Q$ and the coupling constant $\alpha_2$ for the 3-horizon black hole. For negative coupling constant $\alpha_2$, the spacetime can have three horizons. When the coupling constant $\alpha_2$ vanishes, it reduces to the RN spacetime. What is more, the charge $Q$ can be much larger than the mass $M$ while the central singularity remains dressed by an event horizon. \emph{For black holes with $N+1$ horizons, there are $N+1$ physical parameters, the mass $M$, charge $Q$ and the coupling constants $\alpha_i$ with $i$ running over from $2$ to $N$. In this sense, the black holes have $N+1$ ``hairs"}.

We find that, \emph{for the multi-horizon black holes,} both the null geodesics and the timelike geodesics are considerable rich.  For null geodesics, there are: (1) the hyperbolic orbits; (2) the oscillating orbits; (3) the plunging orbits; (4) the stable circular orbit; (5) the unstable circular orbit; (6) the hyperbolic and plunging coalescing orbits; (7) the hyperbolic and the oscillating coalescing orbits; (8) the plunging and the oscillating coalescing orbits; (9) the hyperbolic, the oscillating and the plunging coalescing orbits.
It is found there is one stable circular orbit for the null geodesics. The ratio of radii for null ISCO to event horizon is $\epsilon\simeq1.5055$ which is larger than the Schwarzschild black hole (with $\epsilon\simeq1.5$). Above conclusion are also applicable to timelike geodesics when $E^2>1$. When $E^2<1$ (the bound orbits), the structure of timelike geodesics is richer than that of null geodesics. For example, one can have two oscillating orbits in case (1) and two stable circular orbits in case (7). The ratio of radii for timelike ISCO to event horizon is $\epsilon\simeq2.0678$ which is smaller than the Schwarzschild black hole (with $\epsilon\simeq3$).

For the RN black hole, the Penrose process occurs if and only if the charge of test particle $q$ has different sign from the black hole charge $Q$. However, it is not the case for 3-horizon black hole. It is found the Penrose process can also occur when the two charges, $q$ and $Q$ have the same signs provided that $\alpha_2$ is sufficiently negative.\emph{ This is an interesting property for multi-horizon black holes.} We also calculate the Love numbers of RN black holes and 3-horizon black holes. We find they are all vanishing.
This reveals both the RN black holes and the 3-horizon black holes are rigid. It is the same as the Schwarzschild black holes. Finally, the thermodynamics for 3-horizon black holes is developed. It is found that the inverse of coupling constant $\alpha_2$ plays the role of thermal pressure.

\section*{ACKNOWLEDGMENTS}

This work is partially supported by the Strategic Priority Research Program ``Multi-wavelength Gravitational Wave Universe'' of the
CAS, Grant No. XDB23040100 and the NSFC under grants 11633004, 11773031.


\newcommand\ARNPS[3]{~Ann. Rev. Nucl. Part. Sci.{\bf ~#1}, #2~ (#3)}
\newcommand\AL[3]{~Astron. Lett.{\bf ~#1}, #2~ (#3)}
\newcommand\AP[3]{~Astropart. Phys.{\bf ~#1}, #2~ (#3)}
\newcommand\AJ[3]{~Astron. J.{\bf ~#1}, #2~(#3)}
\newcommand\GC[3]{~Grav. Cosmol.{\bf ~#1}, #2~(#3)}
\newcommand\APJ[3]{~Astrophys. J.{\bf ~#1}, #2~ (#3)}
\newcommand\APJL[3]{~Astrophys. J. Lett. {\bf ~#1}, L#2~(#3)}
\newcommand\APJS[3]{~Astrophys. J. Suppl. Ser.{\bf ~#1}, #2~(#3)}
\newcommand\JHEP[3]{~JHEP.{\bf ~#1}, #2~(#3)}
\newcommand\JMP[3]{~J. Math. Phys. {\bf ~#1}, #2~(#3)}
\newcommand\JCAP[3]{~JCAP {\bf ~#1}, #2~ (#3)}
\newcommand\LRR[3]{~Living Rev. Relativity. {\bf ~#1}, #2~ (#3)}
\newcommand\MNRAS[3]{~Mon. Not. R. Astron. Soc.{\bf ~#1}, #2~(#3)}
\newcommand\MNRASL[3]{~Mon. Not. R. Astron. Soc.{\bf ~#1}, L#2~(#3)}
\newcommand\NPB[3]{~Nucl. Phys. B{\bf ~#1}, #2~(#3)}
\newcommand\CMP[3]{~Comm. Math. Phys.{\bf ~#1}, #2~(#3)}
\newcommand\CQG[3]{~Class. Quantum Grav.{\bf ~#1}, #2~(#3)}
\newcommand\PLB[3]{~Phys. Lett. B{\bf ~#1}, #2~(#3)}
\newcommand\PRL[3]{~Phys. Rev. Lett.{\bf ~#1}, #2~(#3)}
\newcommand\PR[3]{~Phys. Rep.{\bf ~#1}, #2~(#3)}
\newcommand\PRd[3]{~Phys. Rev.{\bf ~#1}, #2~(#3)}
\newcommand\PRD[3]{~Phys. Rev. D{\bf ~#1}, #2~(#3)}
\newcommand\RMP[3]{~Rev. Mod. Phys.{\bf ~#1}, #2~(#3)}
\newcommand\SJNP[3]{~Sov. J. Nucl. Phys.{\bf ~#1}, #2~(#3)}
\newcommand\ZPC[3]{~Z. Phys. C{\bf ~#1}, #2~(#3)}
\newcommand\IJGMP[3]{~Int. J. Geom. Meth. Mod. Phys.{\bf ~#1}, #2~(#3)}
\newcommand\IJMPD[3]{~Int. J. Mod. Phys. D{\bf ~#1}, #2~(#3)}
\newcommand\IJMPA[3]{~Int. J. Mod. Phys. A{\bf ~#1}, #2~(#3)}
\newcommand\GRG[3]{~Gen. Rel. Grav.{\bf ~#1}, #2~(#3)}
\newcommand\EPJC[3]{~Eur. Phys. J. C{\bf ~#1}, #2~(#3)}
\newcommand\PRSLA[3]{~Proc. Roy. Soc. Lond. A {\bf ~#1}, #2~(#3)}
\newcommand\AHEP[3]{~Adv. High Energy Phys.{\bf ~#1}, #2~(#3)}
\newcommand\Pramana[3]{~Pramana.{\bf ~#1}, #2~(#3)}
\newcommand\PTP[3]{~Prog. Theor. Phys{\bf ~#1}, #2~(#3)}
\newcommand\APPS[3]{~Acta Phys. Polon. Supp.{\bf ~#1}, #2~(#3)}
\newcommand\ANP[3]{~Annals Phys.{\bf ~#1}, #2~(#3)}
\newcommand\RPP[3]{~Rept. Prog. Phys. {\bf ~#1}, #2~(#3)}
\newcommand\ZP[3]{~Z. Phys. {\bf ~#1}, #2~(#3)}
\newcommand\NCBS[3]{~Nuovo Cimento B Serie {\bf ~#1}, #2~(#3)}
\newcommand\AAP[3]{~Astron. Astrophys.{\bf ~#1}, #2~(#3)}
\newcommand\MPLA[3]{~Mod. Phys. Lett. A.{\bf ~#1}, #2~(#3)}

\end{document}